\documentclass[epj]{webofc}
\usepackage[utf8]{inputenc}
\usepackage[varg]{txfonts}   
\usepackage{booktabs}
\usepackage{xcolor}
\definecolor{darkred}{rgb}{0.4,0.0,0.0}
\definecolor{darkgreen}{rgb}{0.0,0.4,0.0}
\definecolor{darkblue}{rgb}{0.0,0.0,0.4}
\usepackage[bookmarks,linktocpage,colorlinks,
    linkcolor = darkred,
    urlcolor  = darkblue,
    citecolor = darkgreen]{hyperref}
\usepackage{subfigure}
\wocname{EPJ Web of Conferences}
\woctitle{Lattice2017}
\begin{document}
%
\selectlanguage{english}
\title{%
Topological susceptibility in 2+1-flavor QCD with chiral fermions
}
\author{%
\firstname{Sinya} \lastname{Aoki}\inst{1}\fnsep \and
\firstname{Guido} \lastname{Cossu}\inst{2} \and
\firstname{Hidenori}  \lastname{Fukaya}\inst{3}\fnsep\thanks{Speaker, \email{hfukaya[at]het.phys.sci.osaka-u.ac.jp}} \and
\firstname{Shoji} \lastname{Hashimoto}\inst{4,5} \and
\firstname{Takashi} \lastname{Kaneko}\inst{4,5}
}
\institute{%
Center for Gravitational Physics, Yukawa Institute for Theoretical Physics, Kyoto 606-8502, JAPAN
\and
School of Physics and Astronomy, The University of Edinburgh, Edinburgh EH9 3JZ, United Kingdom
\and
Department of Physics, Osaka University,
Toyonaka 560-0043, Japan
\and
KEK Theory Center,
  High Energy Accelerator Research Organization (KEK),
  Tsukuba 305-0801, Japan
  \and
   School of High Energy Accelerator Science,
  The Graduate University for Advanced Studies (Sokendai),
  Tsukuba 305-0801, Japan
}
\abstract{%
We compute the topological susceptibility $\chi_t$ of 2+1-flavor
lattice QCD with dynamical M\"obius domain-wall fermions,
whose residual mass is kept at 1 MeV or smaller.
In our analysis, we focus on the fluctuation of the topological charge
density in a ``slab'' sub-volume of
the simulated lattice, as proposed by Bietenholz et al.
The quark mass dependence of our results agrees well with the
prediction of the chiral perturbation theory,
from which the chiral condensate is extracted.
Combining the results for the pion mass $M_\pi$ and decay constant $F_\pi$,
we obtain $\chi_t$ = 0.227(02)(11)$M_\pi^2 F_\pi^2$ at the physical point,
where the first error is statistical and the second is systematic.
}
\maketitle
\section{Introduction}\label{intro}

Computing the topological susceptibility, $\chi_t$,
has been a challenging task for lattice QCD.
One of the problems is that the topological charge is sensitive to 
the violation of chiral symmetry.
The quark mass dependence of $\chi_t$
comes entirely from sea quarks, or a small quantum effect suppressed by $O(\hbar)$,
to which contribution from the discretization systematics is relatively large.
Simulating QCD on a sufficiently fine lattice is another challenge,
as the global topological charge tends to be frozen along the Monte Carlo history \cite{Schaefer:2010hu}.
Due to these difficulties, 
the study of the quark mass dependence of $\chi_t$ 
and its comparison with the ChPT formula \cite{Mao:2009sy,Aoki:2009mx,Guo:2015oxa},
\begin{eqnarray}
\label{eq:ChPT}
\chi_t = \frac{m_{ud}\Sigma}{2}\left\{1-\frac{3m_{ud}\Sigma}{16\pi^2F_{\rm phys}^4}\ln \left(\frac{2m_{ud}\Sigma}{F_{\rm phys}^2M_{\rm phys}^2}\right)+\frac{4m_{ud}\Sigma}{F_{\rm phys}^4}l\right\},
\end{eqnarray}
has been very limited, and only some pilot works with dynamical chiral 
fermions on rather small or coarse lattices 
have been performed 
\cite{Aoki:2007pw,Chiu:2008kt,Hsieh:2009zz}.
Here $\Sigma$ denotes the chiral condensate, $l=l_3^r-l_7^r+h_1^r-h_3^r$ 
is a combination of the low-energy constants (renormalized at $M_{\rm phys}$)
at next-to-leading order (NLO)
\cite{Gasser:1983yg}, 
and $M_{\rm phys}$ and $F_{\rm phys}$ are the physical values 
of the pion mass and decay constant, respectively.

In this work 
we improve the computation of $\chi_t$ in three ways.
The first is to employ the
M\"obius domain-wall fermion \cite{Brower:2004xi}
for the dynamical quarks. 
It realizes good chiral symmetry, {\it i.e.} the residual mass is kept at the order of $1$ MeV or lower.
As will be shown below, our results have only a mild dependence on
the lattice spacing, up to $a\sim 0.08$ fm.
Unlike the overlap fermion that we employed in the previous studies \cite{Aoki:2007pw},
the use of the domain-wall fermion allows us to sample configurations in different
topological sectors. 

The second improvement comes from the use of sub-volumes of the simulated lattices.
Since the correlation length of QCD is limited, at most by $1/M_\pi$,
there is no fundamental reason to use the global topological charge to compute $\chi_t$.
The use of sub-volumes was tested in our previous simulations with
overlap quarks \cite{Aoki:2007pw, Hsieh:2009zz},
where
the signal of $\chi_t$ was extracted from finite volume effects
on the $\eta^\prime$ meson correlator \cite{Brower:2003yx, Aoki:2007ka}.
In this work, we utilize a different method,
which was originally proposed by Bietenholz {\it et al.}
\cite{Bietenholz:2015rsa, Bietenholz:2016szu, Mejia-Diaz:2017hhp}
(similar methods were proposed in \cite{Shuryak:1994rr} and \cite{deForcrand:1998ng}).
The method is based on a correlator of local topological charge density,
which is designed to be always non negative, and thus to reduce statisical noise.
We confirm that 30\%--50\% sub-volumes 
of the whole lattice, whose size is $\sim 2$ fm,  
are sufficient to extract $\chi_t$.
Moreover, the new definition shows more frequent fluctuation 
than that of the global topological charge on our finest lattice.

The third improvement is to take the ratio of
the topological susceptibility to $M_\pi^2 F_\pi^2$,
a product of the pion mass $M_\pi$ and decay constant $F_\pi$ squared calculated on the same ensemble.
This ratio,
\begin{eqnarray}
\label{eq:ratio}
\frac{\chi_t}{M_\pi^2F_\pi^2}=\frac{1}{4}\left[1+\frac{2M_\pi^2 l^{\prime}}{F_\pi^2}\right],
\end{eqnarray}
where $l^{\prime}=-l_4^r-l_7^r+h_1^r-h_3^r$ is
a combination of the NLO low energy constants,
is free from the chiral logs, as well as possible finite volume effects at NLO.
Moreover, the chiral limit of the ratio, 1/4, is protected from
the strange sea quark effects.
We can, therefore, precisely estimate the topological susceptibility at the physical point
by measuring $\chi_t$, $M_\pi$, and $F_\pi$ at each simulation point.

We also employ the Yang-Mills (YM) gradient flow 
\cite{Luscher:2010iy, Luscher:2011bx, Bonati:2014tqa}
in order to make the global topological charge close to integers,
to remove the UV divergences, and to reduce the statistical noise.
With these improvements, we achieve good enough statistical precision to investigate
the dependence of $\chi_t$ on the sea quark mass.
In fact, the topological susceptibility shows a good agreement with
the ChPT prediction~(\ref{eq:ChPT}).
We determine the value of chiral condensate, as well as $l^{\prime}$,
taking the chiral and continuum limits from formulas (\ref{eq:ChPT}) and (\ref{eq:ratio}).
The same set of data was also used to calculate the $\eta^\prime$ meson
mass \cite{Fukaya:2015ara}, 
which was extracted from the shorter distance region of 
the correlator of the topological charge density.
Further details of this work may be found in \cite{Aoki:2017paw}.

\section{Numerical set-up}\label{sec:setup}

In the numerical simulation of QCD,
we use the Symanzik gauge action and the M\"obius domain-wall
fermion action for gauge ensemble generations \cite{Kaneko:2013jla, Noaki:2014ura, Cossu:2013ola}.
We apply three steps of stout smearing  of the gauge links 
for the computation of the Dirac operator.
Our main runs of $2+1$-flavor lattice QCD simulations are performed
with two different lattice sizes
$L^3\times T=32^3\times 64$ and $48^3\times 96$, for which
we set $\beta$ = 4.17 and 4.35, respectively.
The inverse lattice spacing $1/a$ is estimated to be 2.453(4)~GeV (for $\beta=4.17$) and 3.610(9)~GeV (for $\beta=4.35$),
using the input $\sqrt{t_0}=0.1465$ fm \cite{Borsanyi:2012zs}.
The two lattices share a similar physical size $L\sim 2.6$ fm.
For the quark masses, we choose two values of 
the strange quark mass $m_s$ around its physical point, 
and 3--4 values of the up and down quark masses $m_{ud}$ for each $m_s$.
For our lightest pion mass around 230 MeV
($am_{ud}$ = 0.0035 at $\beta$ = 4.17)
we simulate a larger lattice $48^3\times 96$ with the same set of the parameters to
control the finite volume effects.
We also perform a simulation on a finer lattice 
$64^3\times 128$ (at $\beta=4.47$ [$1/a\sim 4.5$ GeV] and $M_\pi \sim$ 285 MeV).
For each ensemble, 5000 molecular dynamics (MD) time is simulated.
Numerical works are done with the QCD software package IroIro++ \cite{Cossu:2013ola}.

In this setup, we confirm that the violation of the chiral symmetry in
the M\"obius domain-wall fermion formalism
is well under control.
The residual mass is $\sim 1$~MeV \cite{Hashimoto:2014gta}
by choosing the lattice size in the fifth direction $L_5$ = 12 at $\beta$ = 4.17
and less than 0.2 MeV with $L_5$ =  8  at $\beta$ = 4.35 (and 4.47).

On generated configurations, we perform 500--1640 steps of the
YM gradient flow (using the conventional Wilson gauge action) 
with a step-size $a^2\Delta t=$0.01. 
At every 200--400 steps (depending on the parameters)
we store the configuration of the topological charge density and
compute its two-point correlator.

In the following analysis, we measure the 
integrated auto-correlation time of every quantity, 
following the method proposed in \cite{Schaefer:2010hu}.
The statistical error is estimated by the jackknife method (without binning)
multiplied by the square root of auto-correlation time. 

The pion mass and decay constant are computed combining
the pseudoscalar correlators with local and smeared source operators.
Details of the computation are presented in a separate article \cite{Fahy:2015xka}.

\section{Topology fluctuation in a ``slab''}\label{sec:slab}

We use the conventional gluonic definition of the topological charge density
$q^{\rm lat}(x)$, the so-called clover construction \cite{Bruno:2014ova}. 
Since the YM gradient flow smooths the gauge field
in the range of $\sqrt{8t}\sim 0.5$ fm of the lattice,
a simple summation $Q_{\rm lat}=\sum_x q^{\rm lat}(x)$ over all sites gives
values close to integers.

As is well known, the global topological charge $Q_{\rm lat}$ 
suffers from long auto-correlation time in lattice simulations, especially when the lattice spacing is small.
This is true also in our simulations.
At the highest $\beta=4.47$, $Q_{\rm lat}$ drifts very slowly with auto-correlation time of $O(1000)$.



Instead of using the global topological charge $Q_{\rm lat}$, 
we extract the topological susceptibility from 
a sub-volume $V_{\rm sub}$ of the whole lattice $V$.
Since the correlation length of QCD is limited by at most $1/M_\pi$,
the subvolume $V_{\rm sub}$ should contain sufficient information
to extract $\chi_t$, provided that its size is larger than $1/M_\pi$.
One can then effectively increase the statistics by
$V/V_{\rm sub}$, since each piece of $V/V_{\rm sub}$ sub-lattices
may be considered as an uncorrelated sample.
Moreover, there is no potential barrier among topological sectors:
the instantons and anti-instantons freely come in and go out of
the sub-volume, which should make the auto-correlation time
of the observable shorter than that of the global topological charge.

There are various ways of cutting the whole lattice into sub-volumes
and computing the correlation functions in them.
After some trial and error, we find that the so-called
``slab'' method, proposed by Bietenholz {\it et al.} \cite{Bietenholz:2015rsa}
is efficient for the purpose of computing $\chi_t$.
The idea is to sum up the two-point correlators of the topological charge density,
over $x$ and $y$ in the same sub-volume:
\begin{eqnarray}
\langle Q_{\rm slab}^2(T_{\rm cut}) \rangle &\equiv& \int_{T_{\rm ref}}^{T_{\rm cut}+T_{\rm ref}} 
dx_0 \int_{T_{\rm ref}}^{T_{\rm cut}+T_{\rm ref}} 
dy_0\int d^3x \int d^3y\left\langle q^{\rm lat}(x)q^{\rm lat}(y)\right\rangle.
\end{eqnarray}
Here the integration over $x$ and $y$ in the spatial directions runs in the whole spatial volume
while the temporal sum is restricted to the region 
$[T_{\rm ref}, T_{\rm cut}+T_{\rm ref}]$, which is called a ``slab''.
Since the YM gradient flow is already performed, there is no divergence from the points of $x=y$. 
$T_{\rm ref}$ denotes an arbitrary reference time.
Due to the translational invariance, the slabs
of the same thickness $T_{\rm cut}$ are equivalent, and one can average over $T_{\rm ref}$.
This method is statistically more stable than the other sub-volume method we applied in
\cite{Aoki:2007pw, Hsieh:2009zz}
since $\langle Q_{\rm slab}^2(T_{\rm cut}) \rangle$ is guaranteed to be always positive.

If we sample large statistics on a large enough lattice volume,
$\langle Q_{\rm slab}^2(T_{\rm cut}) \rangle$ should be a fraction $T_{\rm cut}/T$
of $\chi_t V$. Namely, $\langle Q_{\rm slab}^2(T_{\rm cut}) \rangle$ 
should be a linear function in $T_{\rm cut}$.
Its leading finite volume correction can be estimated
using the formula in \cite{Aoki:2007ka}:
\begin{eqnarray}
\label{eq:linear}
\langle Q_{\rm slab}^2(T_{\rm cut}) \rangle = (\chi_t V)\times \frac{T_{\rm cut}}{T}
+C(1-e^{-m_0 T_{\rm cut}})(1-e^{-m_0 (T-T_{\rm cut})}),
\end{eqnarray}
where $C$ is an unknown constant, and $m_0$ is the mass of the 
first excited state, the $\eta'$ meson.
Note that for $1/m_0 \ll T_{\rm cut}\ll T-1/m_0$, the formula gives a simple 
linear function in $T_{\rm cut}$ plus a constant.
Also, note that in the limit of $T_{\rm cut}=T$, 
$\langle Q_{\rm slab}^2(T_{\rm cut}=T) \rangle$ reduces to the global topology $\langle Q^2\rangle = \chi_t V$.



Assuming the linearity in $T_{\rm cut}$, one
can extract the topological susceptibility through
\begin{eqnarray}
\label{eq:chitslab1}
\chi_t^{\rm slab} = \frac{T}{V}\left[\frac{\langle Q_{\rm slab}^2(t_1) \rangle - \langle Q_{\rm slab}^2(t_2) \rangle}{t_1-t_2}\right],
\end{eqnarray}
with two reference thicknesses $t_1$ and $t_2$.
In our numerical analysis, $T_{\rm ref}$ is averaged over the temporal direction.
Since the data at $t_i$ and $T-t_i$ are not independent,
we choose $t_1$ and $t_2$ in a range 1.6 fm $<t_1, t_2< T/2$. 
In the numerical analysis,
we replace $q^{\rm lat}(x)$ by $q^{\rm lat}(x)-\langle Q_{\rm lat}/V\rangle$
to cancel a possible bias due to the long auto-correlation of the global topology.


We find that the signal using this slab method
is less noisy than the previous attempts
in \cite{Aoki:2007pw, Hsieh:2009zz}.
Moreover,
the new method shows more frequent fluctuation 
than that of the global topological charge on our finest lattice.


%
\section{Results}
\label{sec:results}

At $\beta=4.17$, which corresponds to the lattice spacing $a\sim 0.08$ fm,
both $Q_{\rm lat}$ and $Q_{\rm slab}^2(T_{\rm cut})$
fluctuate well.
The data on this lattice, therefore, provide a good testing ground to 
examine the validity of the slab sub-volume method, comparing 
with the naive definition of the topological
susceptibility, $\langle Q_{\rm lat}^2\rangle/V$.

\begin{figure*}[thb]
  \centering
 \includegraphics[width=10cm]{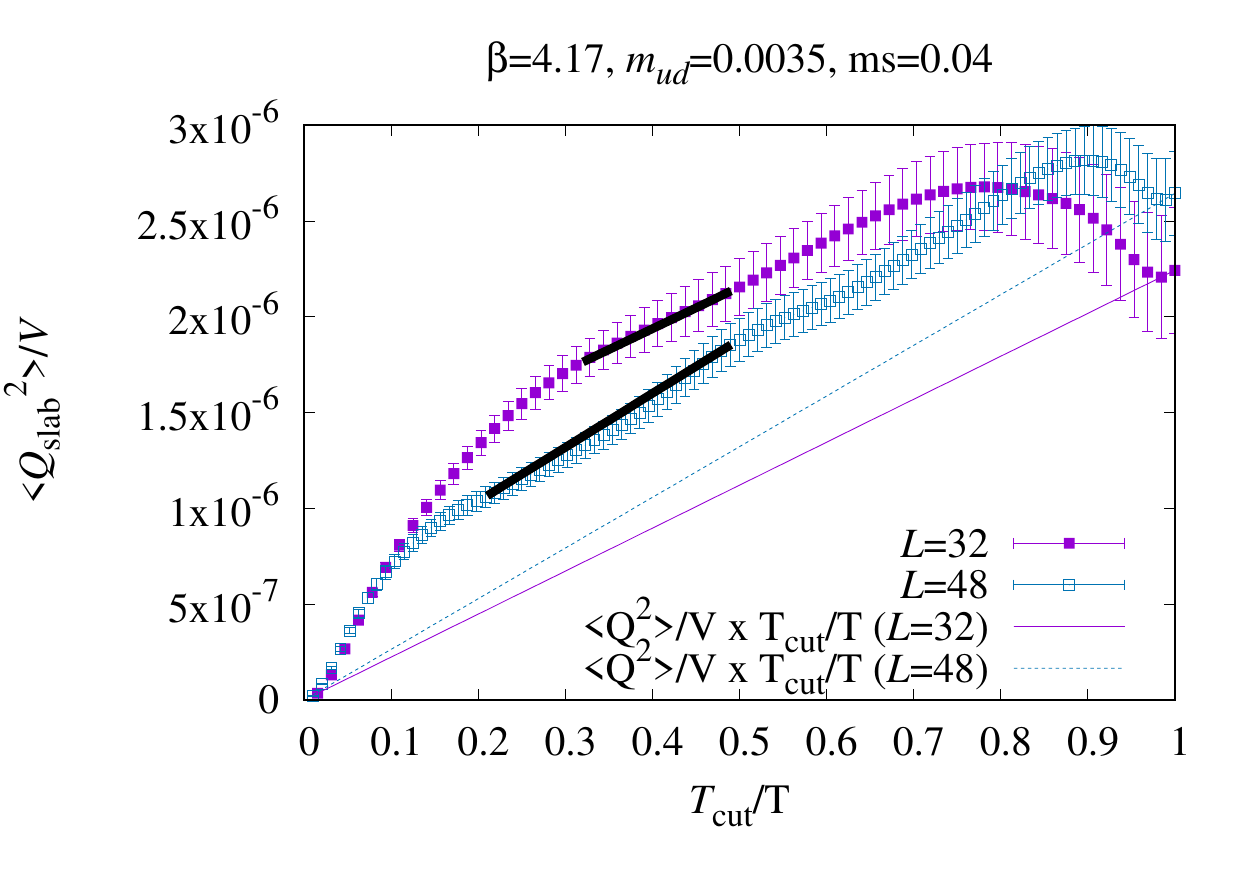}
\caption{
 $\langle Q_{\rm slab}^2(T_{\rm cut})\rangle$ as a function of $T_{\rm cut}/T$.
  Data at lightest mass $m_{ud}=0.0035$, $\beta=4.17$ with two different lattice sizes
  $L=32$ and $L=48$ are shown. $T=2L$ for both lattices.
 Two end-points of the thick line segments show the reference points $t_1$ and $t_2$
 taken for determination of the topological susceptibility.
 Note that the value of $t_1=20$ is the same for the two data.
}
\label{fig:finiteV2}
\end{figure*}

In Fig.~\ref{fig:finiteV2}, $\langle Q_{\rm slab}^2(t_{\rm cut}) \rangle$
observed at the lightest sea quark mass $m_{ud}=0.0035$, $\beta=4.17$
on two different volumes $L=32$ and $L=48$ are plotted as a function of $T_{\rm cut}/T$.
The data converge to the linear plus constant function~(\ref{eq:linear}) at $T_{\rm cut}=20$, which corresponds to $\sim 1.6$ fm.
 The slope, or $\chi_t^{\rm slab}$, is consistent with that from global topology 
 shown by solid and dotted lines for $L=32$ and $L=48$ lattices, respectively.  
 We also observe the consistency between the $L=32$ and $L=48$ data,
 which suggests that the systematics due to the finite volume is well under control.
The ``linear $+$ constant'' behavior is also seen in ensembles with heavier quark masses.
Moreover, the extracted values of the topological susceptibility from the slope
show a good agreement with the ChPT prediction,
which confirms the validity of the slab method.

\begin{figure*}[thb]
  \centering
 \includegraphics[width=8cm]{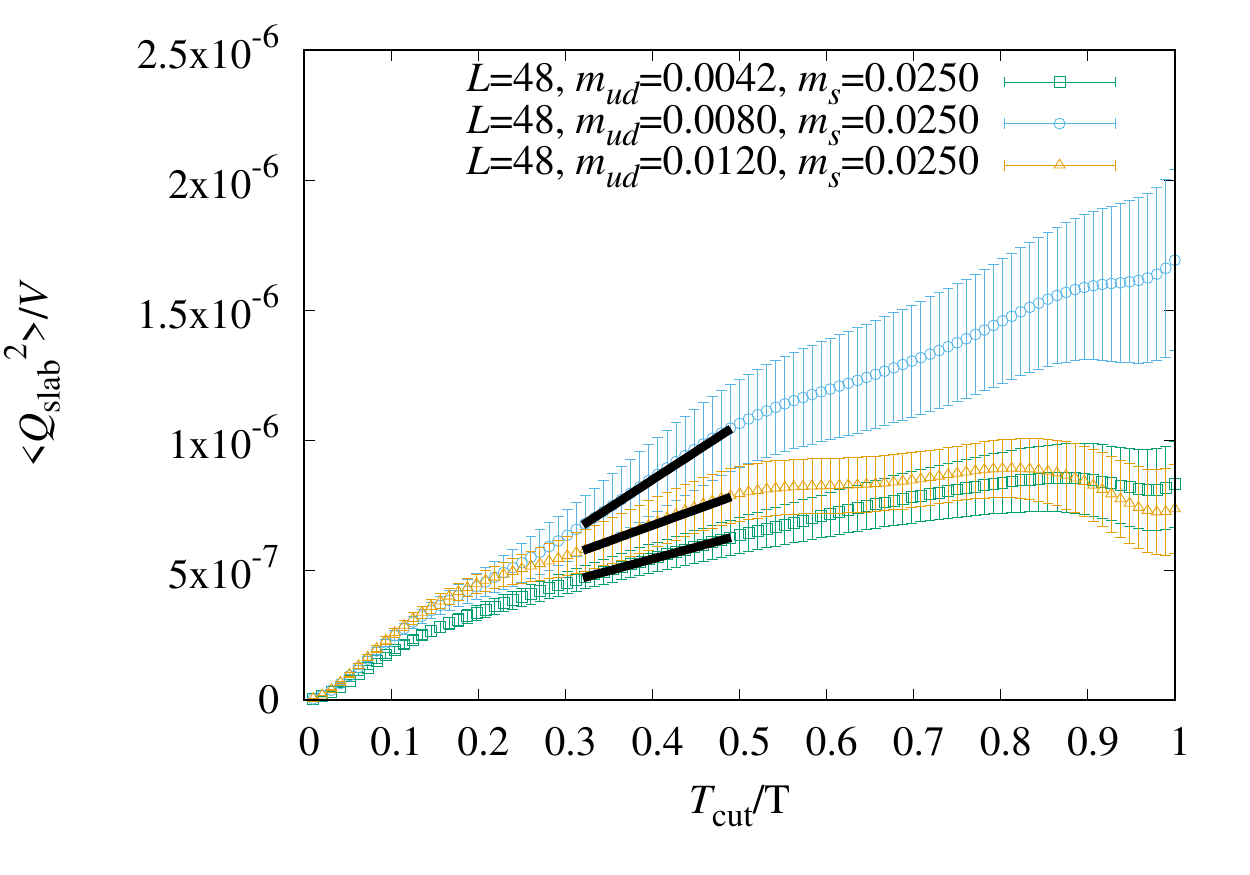}
\caption{
  $\langle Q_{\rm slab}^2(T_{\rm cut})\rangle$ at different up and down quark masses.
  Data at $\beta=4.35$ and $m_s=0.0180$ are shown.
}
\label{fig:chithighbeta}
\end{figure*}

At higher $\beta$ values, we also find a reasonable
slope at the lightest quark mass for each $\beta$ and $m_s$,
as shown in Fig.~\ref{fig:chithighbeta}.
For heavier masses, however, some curvature is seen.
We consider this curvature is an effect from the bias of the global topological charge.
This observation is consistent with previous works (see \cite{Schaefer:2010hu} for example),
which reported that the heavier pion mass ensembles
show the longer auto-correlation of the topological charge,
and the larger deviation of $\langle Q_{\rm lat}\rangle$ from zero.
We determine the shorter reference $t_1\sim 1.6$ fm
using the data at the lightest quark mass and always choose $t_2=T/2 \sim 2.6$ fm.
In order to estimate the systematic errors due to non-linear behavior,
we compare the results 
with 1) those obtained from different reference times $(t_1', t_2')=(t_1,\frac{t_1+t_2}{2})$, and $(\frac{t_1+t_2}{2},t_2)$,
and 2) those obtained without the subtraction of $\langle Q\rangle/V$ in the definition of the topological charge density.
The larger deviation is treated as a systematic error.
For the statistical error estimates we follow the method proposed by
the ALPHA collaboration \cite{Schaefer:2010hu}
assuming a double exponential structure of the autocorrelation function.

\begin{figure*}[thb]
  \centering
  \includegraphics[width=8cm]{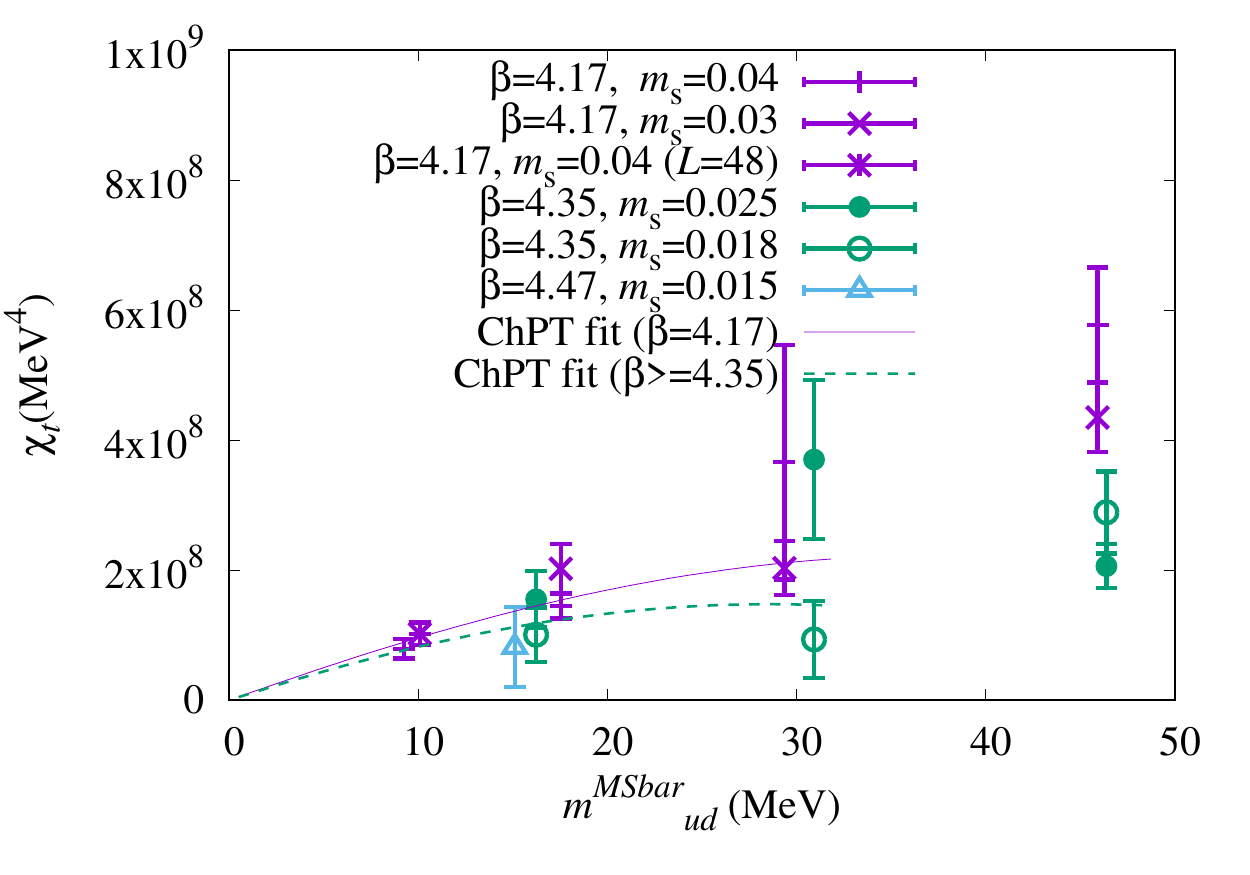}
 \includegraphics[width=8cm]{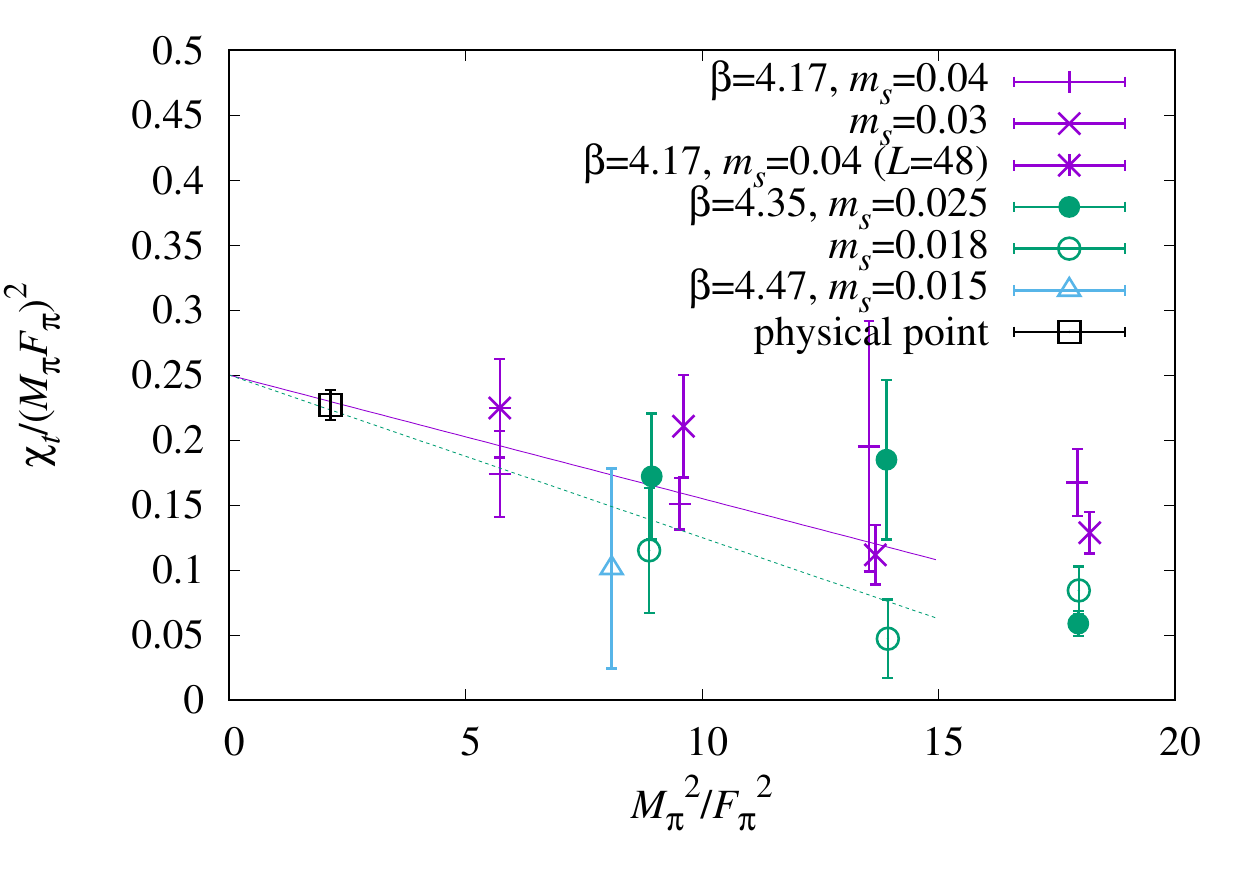}
\caption{
  $m_{ud}$ dependence of topological susceptibility (top)
  and the ratio $\frac{\chi_t}{M_\pi^2F_\pi^2}$ (bottom) obtained from
  the slab sub-volume method.
  The heaviest four points are not included in the fit.
}
\label{fig:result}
\end{figure*}

The top panel of Fig.~\ref{fig:result} presents 
our data for $\chi_t^{\rm slab}$ from all ensembles
plotted in physical units.
The error-bar in the plot represents
the statistical and systematic errors
due to the choice of the references $t_1$ and $t_2$, added in quadrature.
On the horizontal axis, the quark mass defined
in the $\overline{\mbox{MS}}$ scheme at 2 GeV is 
\begin{eqnarray}
  m_{ud}^{\scriptsize \overline{\mbox{MS}}} = (m_{ud}+m_{res})/Z_S,
\end{eqnarray}
where the renormalization factor $Z_S$ is nonperturbatively computed
in \cite{Tomii:2016xiv}.
We find a linear suppression near the chiral limit.
We also find that there is no clear
dependence on $\beta$ and $m_s$.

First, we compare our results directly to the ChPT formula (\ref{eq:ChPT}).
We perform a two-parameter ($\Sigma$ and $l$) fit
to the data at $\beta=4.17$ (solid curve in Fig.~\ref{fig:result})
and $\beta \ge 4.35$ (dashed curve) separately.
Here we also perform the same fit but omitting the heaviest two points,
and take the difference as an estimate for the systematic error
in the chiral extrapolation.
Since the heaviest points have the following problems,
1) a strong bias is seen in the global topology,
2) ChPT is less reliable, and 3)~there is a mismatch between different $\beta$,
we take the result without them as our central values.
Note, however, this inclusion/elimination affects $l$ but $\Sigma$ is stable against
the change in the fit-range.
Namely, the chiral condensate $\Sigma$ is effectively determined by the lower quark mass data.
We then estimate the continuum limit by a constant fit.
Comparing our result from the constant fit with a linear
extrapolation of the central values,
we take the difference as an estimate for the systematic error
in the continuum extrapolation.

Next, using our data for the pion mass $M_\pi$ and decay constant $F_\pi$
together with $\chi_t^{\rm slab}$, 
obtained from each ensemble, we take the ratio~(\ref{eq:ratio}).
By a linear one-parameter fit as shown in the bottom panel of Fig.~\ref{fig:result},
we determine $l^\prime$ and the ratio $\chi_t^{\rm slab}/(M_\pi F_\pi)^2$ at the physical point.
In the same way as the determination of $\Sigma$ and $l$,
we take the chiral and continuum limits of both quantities.
Note that the fixed chiral limit at $1/4$ of the ratio helps us 
to determine these quantities.

Finally let us discuss other possible systematic effects.
In our analysis, the ensembles satisfying $M_\pi L>3.9$ are used 
and  we do not expect any sizable finite volume effects.
In particular, our lightest mass point has $M_\pi L=4.4$.
We have used configurations at the YM gradient flow-time
around $\sqrt{8t}\sim 0.5$~fm.
We confirm that the flow--time dependence is negligible in the range
$0.25$ fm $< \sqrt{8t} < 0.5$~fm.
We conclude that all these systematic effects are 
negligibly small compared to the statistical and systematic errors given above.

\section{Summary and discussion}

With dynamical M\"obius domain-wall fermions and the new method
using slab sub-volumes of the simulated lattice,
we have computed the topological susceptibility of QCD.
Its quark mass dependence is consistent with
the ChPT prediction, from which we have obtained
\begin{eqnarray}
\chi_t &=& 0.227(02)(03)(11) M_\pi^2 F_\pi^2\;(\mbox{at physical point}),\\
\Sigma^{\overline{\rm MS}}(\mbox{2GeV}) &=& [270(12)(21)(07)\mbox{MeV}]^3,
\end{eqnarray}
where the first error comes from the statistical uncertainty
at each simulation point, including the effect of freezing topology.
The second and third represent the systematics in the chiral and continuum limits, respectively.
The value of $\Sigma$ is consistent with our recent determination through the Dirac spectrum \cite{Cossu:2016eqs}.
We have also estimated the NLO coefficient
\begin{eqnarray}
l &=&(l_3^r-l_7^r+h_1^r-h_3^r)=-0.002(03)(05)(15),\\
l^\prime &=& (-l_4^r-l_7^r+h_1^r-h_3^r)= -0.021(02)(02)(10),
\end{eqnarray}
where $l$ is renormalized at the physical pion mass, while $l^\prime$ is renormalization invariant.
It is interesting to note that $l$ and $l^\prime$ include
a combination of the coefficients $h_1^r-h_3^r$, 
which is supposed to be {\it unphysical}
in ChPT unless $\theta$ dependence is considered.
These are important for possible
couplings of QCD to axions \cite{diCortona:2015ldu}.

\vspace*{5mm}

\vspace*{5mm}
We thank T.~Izubuchi, and other 
members of JLQCD collaboration for useful discussions.
Numerical simulations are performed on IBM System Blue Gene Solution at KEK under 
a support of its Large Scale Simulation Program (No. 16/17-14). 
This work is supported in part 
by the Japanese Grant-in-Aid for Scientific Research
(Nos. JP25800147, JP26247043, JP26400259, JP16H03978), 
and by MEXT as ``Priority Issue on Post-K computer''
(Elucidation of the Fundamental Laws and Evolution of
the Universe) and by Joint Institute for Computational Fundamental Science (JICFuS).
The work of GC is supported by STFC, grant ST/L000458/1.


\end{document}